\documentclass[prb,aps,superscriptaddress,reprint]{revtex4-1}

\usepackage{graphicx}% Include figure files
\usepackage{dcolumn}% Align table columns on decimal point
\usepackage{bm}% bold math
\usepackage{setspace}
\usepackage{amsmath}
\usepackage{epsfig}

\newcommand{\vtm}{$\mathrm{V_{TM}}$}
\newcommand{\vtv}{$\mathrm{V_{TV}}$}
\newcommand{\itm}{$\mathrm{I_{TM}}$}
\newcommand{\itv}{$\mathrm{I_{TV}}$}

\begin{document}

\title{Localised Solid-State Nanopore Fabrication via Controlled Breakdown using On-Chip Electrodes}

\author{Jasper P.~Fried}
\affiliation{Department of Materials, University of Oxford, Oxford, OX1 3PH, U.K.}
\affiliation{School of Chemistry, University of
New South Wales, Sydney, New South Wales 2052, Australia}
\email{j.fried@unsw.edu.au}
\author{Jacob L.~Swett}
\affiliation{Department of Materials, University of Oxford, Oxford, OX1 3PH, U.K.}
\author{Binoy Paulose Nadappuram}
\author{Aleksandra Fedosyuk}
\affiliation{Department of Chemistry, Imperial College London, London, W12 0BZ, U.K.}
\author{Alex Gee}
\affiliation{Department of Materials, University of Oxford, Oxford, OX1 3PH, U.K.}
\author{Ondrej E.~Dyck}
\affiliation{Center for Nanophase Materials Sciences, Oak Ridge National Laboratory, Oak Ridge, TN, 37830 U.S.A.}
\author{James R. Yates}
\affiliation{Instituto de Tecnologia Química e Biológica António Xavier
Universidade Nova de Lisboa, Av. da República, Oeiras 2780-157, Portugal}
\author{Aleksandar P. Ivanov}
\author{Joshua B. Edel}
\affiliation{Department of Chemistry, Imperial College London, London, W12 0BZ, U.K.}
\author{Jan A.~Mol}
\affiliation{School of Physics and Astronomy, Queen Mary University of London, London, E1 4NS, U.K.}

\date{\today}% It is always \today, today,

%%%%%%%%%%%%%%%% ABSTRACT %%%%%%%%%%%%%%%%

\pacs{}

\begin{abstract}

Controlled breakdown has recently emerged as a highly accessible technique to fabricate solid-state nanopores. However, in its most common form, controlled breakdown creates a single nanopore at an arbitrary location in the membrane. Here, we introduce a new strategy whereby breakdown is performed by applying the electric field between an on-chip electrode and an electrolyte solution in contact with the opposite side of the membrane. We demonstrate two advantages of this method. First, we can independently fabricate multiple nanopores at given positions in the membrane by localising the applied field to the electrode. Second, we show we can create nanopores that are self-aligned with complementary nanoelectrodes by applying voltages to the on-chip electrodes to locally heat the membrane during controlled breakdown. This new controlled breakdown method provides a path towards the affordable, rapid, and automatable fabrication of arrays of nanopores self-aligned with complementary on-chip nanostructures.  

\end{abstract}

\maketitle

Solid-state nanopore devices have received interest in recent years for applications ranging from protein analysis \cite{Yusko2016,Alfaro2021,Restrepo-Perez2018}, to polymer data storage \cite{Bell2016,Chen2020}, and ultra-dilute analyte detection \cite{Chuah2019,Wu2020,Cai2021,Rozevsky2020}. These devices typically consist of a nanometre-scale hole in a synthetic material that separates two chambers of electrolyte solution. When a voltage is applied across the membrane, ions flow through the nanopore resulting in a measurable ionic current. Sensing is then typically achieved by detecting changes in the ionic current as an analyte translocates through the pore and modifies the flow of ions \cite{Dekker2007,Miles2013,Xue2020}. 

In recent years, there has been a significant interest in integrating solid-state nanopores with on-chip electrodes. In particular, electrodes embedded within, or in close proximity to a nanopore can be used to control the translocation dynamics of biomolecules \cite{Tsutsui2021,Luan2010,Paik2012,Nam2009} or enable dielectrophoretic concentrating of analytes at the nanopore opening \cite{Freedman2016}. Solid-state nanopores have also been integrated with field-effect sensors \cite{Xie2011,Traversi2013,Heerema2018,Graf2019,Zhu2021} and tunnelling nanogap electrodes \cite{Ivanov2011,Ivanov2014,Fanget2013,Tsutsui2011,Tang2021}. Measuring the conduction through such nanoelectrodes provides an alternative readout mechanism to ionic current based detection. Integrating a nanopore with a plasmonic nanostructure can also enable/enhance optical nanopore sensing strategies \cite{Cecchini2013, Freedman2016a, Shi2016, Shi2018, Verschueren2018a}. Importantly, unlike ionic current based detection, these alternative readout mechanisms do not require individual nanopores to be fluidically isolated for the signal from each pore to be read out independently \cite{Xie2011}. As such, these alternate sensing modalities can increase the obtainable device density thus enabling high-throughput, parallel detection for quantitative analysis. Moreover, these alternative readout mechanisms can provide sensitivity to molecular properties not possible using ionic current based detection \cite{Chang2009,Zhu2021,Zhang2019} and increase the detection bandwidth \cite{Xie2011,Parkin2018}.  

Despite the advantages provided by integrating on-chip electrodes with solid-state nanopores, the potential of these devices is yet to be fully explored. This has been partly due to challenges associated with accurately and reliably aligning a nanopore with the on-chip nanoelectrodes \cite{Healy2012,Heerema2018,Fried2018}. In the past, this was often done by drilling the nanopore using a focused beam of charged particles such as those created in a transmission electron microscope (TEM). However this has several drawbacks including (i) manual alignment is needed, (ii) it is a low throughput process, (iii) it requires trained operators, and (iv) it utilises expensive equipment. Developing scalable and accessible methods to fabricate solid-state nanopores integrated with on-chip nanoelectrodes would significantly accelerate research into this field. 

In recent years, controlled breakdown (CBD) has emerged as an accessible method to create solid-state nanopores \cite{Kwok2014,Waugh2020,Fried2021}. This technique relies on applying a large electric field across the membrane to induce breakdown and nanopore formation in the dielectric. In its most commonly used form, CBD is performed by applying the breakdown voltage to electrolyte solutions in contact with either side of the membrane. However, this method should only be used to fabricate a single nanopore since new breakdown events will result in the uncontrollable expansion of previously created pores \cite{Waugh2020,Leung2020}. Moreover, the most common form of CBD does not allow for control of the resulting pore position \cite{Zrehen2017}. To overcome this issue, several CBD techniques have been developed that enable control over the nanopore position. These have included using an atomic force microscope tip to apply the voltage \cite{Zhang2019,Yazda2021}, thinning a region of the membrane prior to breakdown \cite{Carlsen2017}, and confining the electrolyte on one side of the membrane \cite{Arcadia2017,Yin2020,Tahvildari2015}. However, to date, none of these techniques have been used to integrate nanopores with on-chip nanostructures. Moreover, some of these methods require relatively expensive equipment thus reducing the main appeal of CBD. Another technique that has been demonstrated relies on illuminating a plasmonic nanostructure with a laser during CBD. This results in nanopore formation at the center of the plasmonic hotspot \cite{Pud2015}. To date, this is the only CBD method that has been used to integrate nanopores with on-chip nanostructures. While this is a highly appealing technique, it can only be used to fabricate a single nanopore in the membrane since additional breakdown events would result in the uncontrollable expansion of previously created pores. Moreover, this method requires optical equipment and can only be used to integrate nanopores with plasmonic nanostructures thus limiting its applicability for certain applications. 

Here, we report a new CBD strategy whereby breakdown is performed by applying a voltage between an on-chip electrode and an electrolyte solution in contact with the other side of the membrane. This new CBD strategy possesses two main advantages. First, by localising the applied electric field to an on-chip electrode we are able to independently fabricate multiple nanopores in the membrane and control their location. Second, by applying appropriate voltages to the on-chip electrodes to locally heat the membrane during CBD, we are able to fabricate nanopores self-aligned with complementary nanostructures. With further development, we are confident that this CBD strategy could be used to fabricate high density arrays of nanopores self-aligned with on-chip nanostructures in a rapid and affordable fashion. The development of such a technique would significantly accelerate research in this field and open up routes for the commercial development of these devices. 

\section{Independent Fabrication of Multiple Nanopores}

\label{sindpores}

\begin{figure*}
    \centering
    \includegraphics[width=\textwidth]{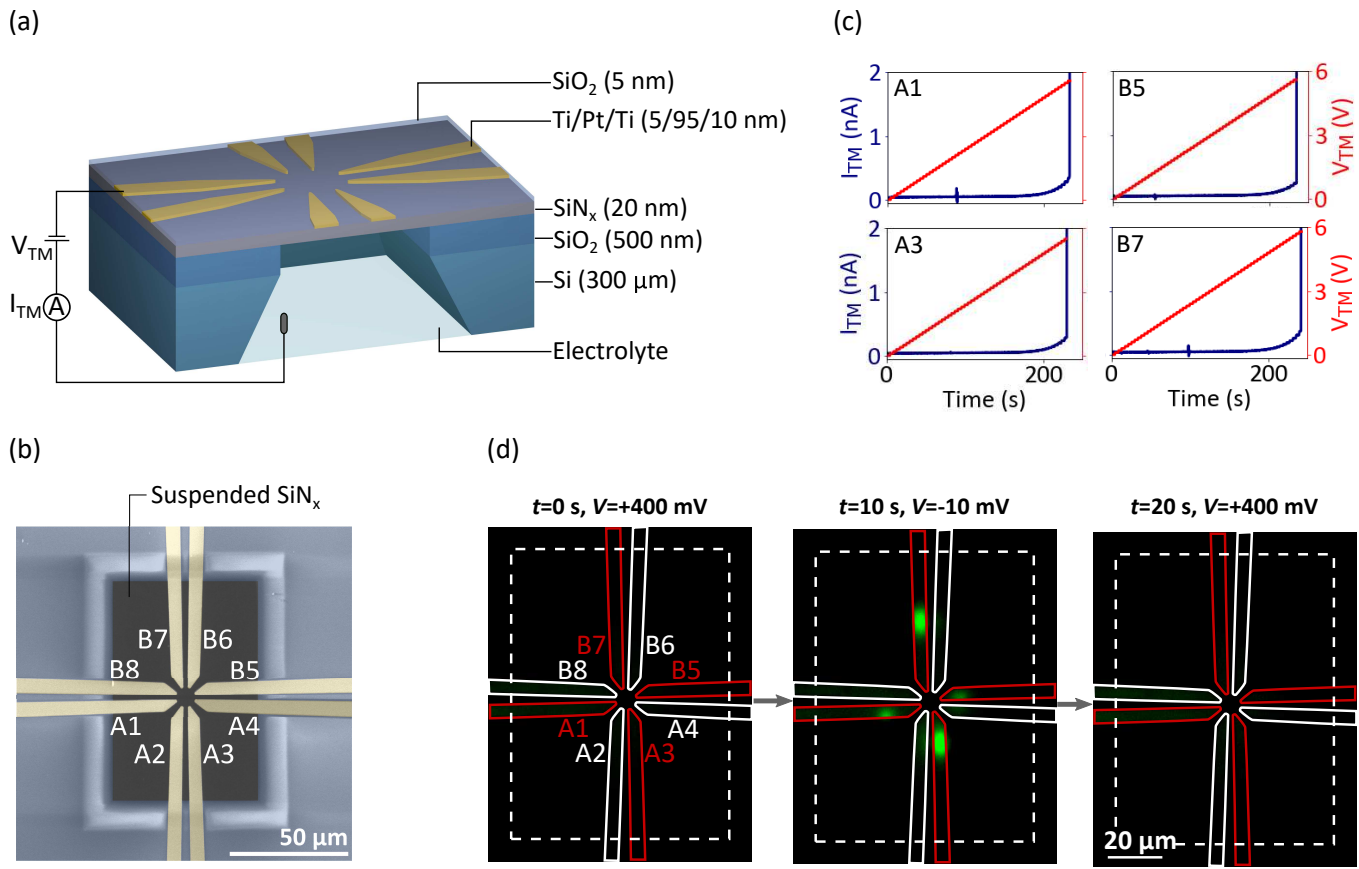}
    \caption{(a) Schematic of the device geometry and experimental setup used to independently fabricate nanopores when the breakdown voltage is applied between an on-chip electrode and an electrolyte solution. (b) False-colour scanning electron micrograph of the electrode configuration over the suspended region of SiN$_x$ (dark grey rectangle). Electrodes are labeled A1-A4 and B5-B8 for convenience in the below discussion (the labels are not part of the device geometry). (c) CBD measurements performed on four different electrodes on the same membrane. The electrode which CBD was performed on is shown in the upper left-hand corner of each plot. (d) Fluorescence micrographs of the nanopores created from the CBD measurements shown in (c). There are three micrographs showing a time series of data with a frame before (left), during (middle), and after (right) the application of a voltage that drives Ca$^{2+}$ ions through the nanopore. The dashed white box shows the edge of the region of suspended SiN$_x$. The solid lines show the electrode positions with red lines for electrodes CBD was performed on and white lines for electrodes CBD was not performed on.}
    \label{findpores}
\end{figure*}

First, we will demonstrate that multiple nanopores can be independently fabricated via CBD by applying the breakdown voltage between an on-chip electrode and an electrolyte solution in contact with the other side of the membrane. A schematic of the device geometry and experimental setup used for these measurements is shown in Fig.~\ref{findpores}(a). A false-colour scanning electron microscope (SEM) image of the electrode configuration over the suspended region of SiN$_x$ is shown in Fig.~\ref{findpores}(b). The devices consist of a thin SiN$_x$ membrane suspended on 500$\,$nm of SiO$_2$ on a Si substrate. There are eight independently addressable metal (5/95/10$\,$nm Ti/Pt/Ti) electrodes patterned on the SiN$_x$ membrane (for convenience we have labeled these A1-A4 and B5-B8 as shown in Fig.~\ref{findpores}(b)). Finally, a thin passivation layer of SiO$_2$ (5$\,$nm) is deposited over the entire membrane surface via atomic layer deposition. A passivation layer is commonly used in nanopore devices integrated with on-chip electrodes to reduce electrochemical reactions between the electrodes and the electrolyte solution \cite{Healy2012}. A more detailed description of the fabrication process is given in SI 1. To enable control of the voltage applied to the on-chip electrodes and the electrolyte solution in contact with the membrane we designed a fluidic cell. This fluidic cell consists of a printed circuit board (PCB) to which the on-chip electrodes are wire-bonded. The PCB is then sandwiched between two PTFE blocks containing perfluoroelastomer O-rings to enable fluidic isolation of the electrolyte solution (described in more detail in SI 2). 

To perform breakdown, a voltage ramp is applied to one of the on-chip electrodes while the electrolyte solution (and the other on-chip electrodes) are held at 0$\,$V (ground). Since the electric field is localised to the electrode, this process can be repeated using different on-chip electrodes to apply the voltage thus enabling independent fabrication of multiple nanopores in the membrane. Figure \ref{findpores}(c) shows CBD measurements performed on four different electrodes on the same membrane. The leakage current prior to breakdown and the breakdown voltage are similar for each measurement, highlighting the ability to independently fabricate multiple nanopores in the same membrane. Note that, due to the asymmetry of this device geometry, the conduction and breakdown characteristics will change significantly depending on the direction of the applied electric field (as explained in Ref.~\cite{Fried2021a}). The results shown here were obtained by applying a negative voltage to the on-chip electrodes and grounding the electrolyte solution. However, we were also able to independently fabricate multiple nanopores in the membrane when the electric field direction was reversed as shown in SI 3. Due to the different breakdown characteristics, it may be necessary to change the voltage protocol required to fabricate nanopores when reversing the electric field direction (particularly the rate at which the voltage is reduced following breakdown). A detailed description of the voltage protocols used for each electric field direction is given in SI 4. 

%This results in breakdown occurring at a relatively low voltage since oxidation reactions does not need to occur at the membrane-electrolyte interface to inject electrons into the membrane (previous studies have demonstrated this is the limiting factor for conduction during CBD \cite{Fried2021a}).

To confirm the independent fabrication of multiple nanopores using this CBD strategy, we have performed fluorescence imaging of the pores. This technique relies on adding Ca$^{2+}$ ions to the solution on one side of the membrane and the Ca$^{2+}$ indicator Fluo-4 to the solution on the other side of the membrane \cite{Anderson2014,Ivankin2014,Zrehen2017}. When an appropriate voltage is applied across the membrane, Ca$^{2+}$ ions are driven through the nanopore resulting in a fluorescence signal at the nanopore. A detailed description of the procedure used for fluorescence imaging of the nanopores is provided in SI 5. Fig.~\ref{findpores}(d) shows fluorescence micrographs of the nanopores created by the CBD measurements shown in Fig.~\ref{findpores}(c). The images represent a time-series of data with a frame before, during, and after the application of a voltage that drives Ca$^{2+}$ ions through the nanopore. The white dashed box represents the suspended region of SiN$_x$. The solid red and white lines show the position of the electrodes that CBD was and was not performed on, respectively. There was a fluorescence signal observed in each electrode on which CBD was performed (differences in the fluorescence intensity correspond to variations in the pore diameters). Conversely no fluorescence signal was observed in electrodes on which CBD was not performed. This result confirms the ability to independently localise and fabricate multiple nanopores in the membrane using this CBD strategy. Note that, it is not trivial that breakdown will result in a nanopore that extends through the electrode and the passivation layer covering the membrane surface. However, the above fluorescence images indicate that the nanopore is created through the entire membrane since a large fluorescence signal will only be observed if the Ca$^{2+}$ binds to the indicator dye. TEM images of the created nanopore (SI 6) and single-molecule biosensing experiments discussed below also confirm that the nanopore extends through the entire membrane.

To obtain statistics of the diameter of the created pores, we have created a single nanopore in the membrane for six different devices using our CBD strategy. After breakdown, reservoirs on both sides of the membrane are filled with electrolyte solution and the ionic current through the nanopore measured to estimate the pore diameter. The CBD protocol used results in the consistent fabrication of nanopores with diameters in the range 1-10$\,$nm (SI 7). One drawback of our CBD strategy described above is that it does not enable direct feedback of the nanopore diameter during fabrication. This is typically done after nanopore creation via CBD by applying voltage pulses to expand the nanopore while recording the ionic current at low voltage ($\sim$200$\,$mV) between each pulse to enable feedback on the pore size \cite{Beamish2012,Waugh2020}. In our CBD setup a current can be measured between the Ag/AgCl electrode and the small area of the on-chip electrode that is exposed to the solution after breakdown. However, this current can not be used to directly estimate the nanopore size \cite{Kowalczyk2011} since there is a non-negligible resistance associated with charge transfer to the on-chip electrode. However, we observe that the current measured while reducing the voltage to zero after breakdown correlates with the resulting nanopore size (SI 7). Namely, measuring a larger current between the Ag/AgCl electrode and the on-chip electrode when reducing the voltage after breakdown generally indicates the formation of a larger nanopore. As such, precise control over the nanopore size may be possible using our CBD strategy by implementing active feedback conditions whereby the voltage is reduced at a rate depending on the measured current.  

\section{Self-Aligning Nanopores with an On-Chip Metal Nanoconstriction}

Above we demonstrated that performing CBD by applying the breakdown voltage between an on-chip electrode and an electrolyte solution in contact with the other side of the membrane enables the independent fabrication of multiple nanopores. To fully exploit this result, it is necessary to utilise alternative sensing mechanisms to ionic current based detection that do not require individual nanopores to be fluidically isolated for their signal to be read out independently. Such alternative readout mechanisms often rely on integrating on-chip nanostructures such as field-effect sensors \cite{Xie2011,Traversi2013,Heerema2018,Graf2019,Zhu2021}, tunnelling nanogaps \cite{Ivanov2011,Ivanov2014,Fanget2013,Tsutsui2011}, plasmonic nanostructures \cite{Spitzberg2019,Garoli2019}, and radiofrequency antennas \cite{Bhat2018} with a nanopore. To address this, we extended our CBD strategy to fabricate nanopores that are self-aligned with on-chip nanoelectrodes. In particular, we show that by locally heating the membrane by applying appropriate voltages to the on-chip electrodes we are able to create nanopores self-aligned with complementary nanoelectrodes. 

\begin{figure}
    \centering
    \includegraphics[width=7cm]{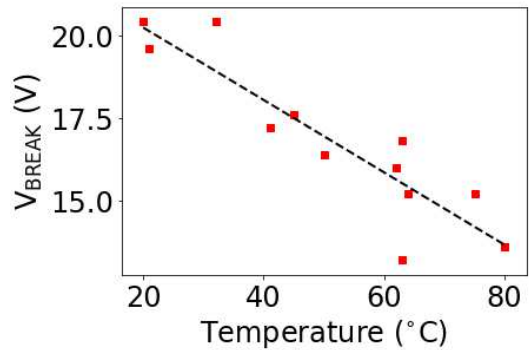}
    \caption{Breakdown voltage as a function of temperature. The dashed black line is a line of best fit which has a gradient of -109$\,\pm$16$\,$mV/$^{\circ}$C (the quoted error is the standard error of the regression slope). This experiment was conducted using the typical CBD configuration where the voltage is applied via electrolyte in contact with both sides of the membrane. Similar results are also observed when the voltage is applied via an on-chip electrode (SI 8).}
    \label{ftempcbd}
\end{figure}

To first understand how heating the membrane affects the breakdown voltage, we have performed CBD under a range of different temperatures. This was done by placing the fluidic cell on a heater plate for 30 minutes for the temperature of the cell to reach equilibrium. The electrolyte reservoirs were covered during this period to avoid evaporation of the solution. Fig.~\ref{ftempcbd} shows the extracted breakdown voltage as a function of temperature. We observe a linear decrease in the breakdown voltage with increasing temperature that is consistent with previous studies \cite{Yanagi2020}. The gradient of the line of best fit (black dashed line) is -109$\,\pm$16$\,$mV/$^{\circ}$C. Note that the measurements shown in Fig.~\ref{ftempcbd} were performed using a typical CBD configuration where electrolyte is present on both sides of the membrane. However, similar results are also observed when the voltage is applied via an on-chip electrode (SI 8). The mechanism for the decrease in breakdown voltage with increasing temperature is not fully understood and may be the result of increased electron transfer processes at the membrane-electrolyte interface or increased electron transport across the dielectric \cite{Fried2021a}. Our results nonetheless indicated that it may be possible to confine nanopore formation to a specific region by locally heating the membrane. 

\begin{figure*}
    \centering
    \includegraphics[width=\textwidth]{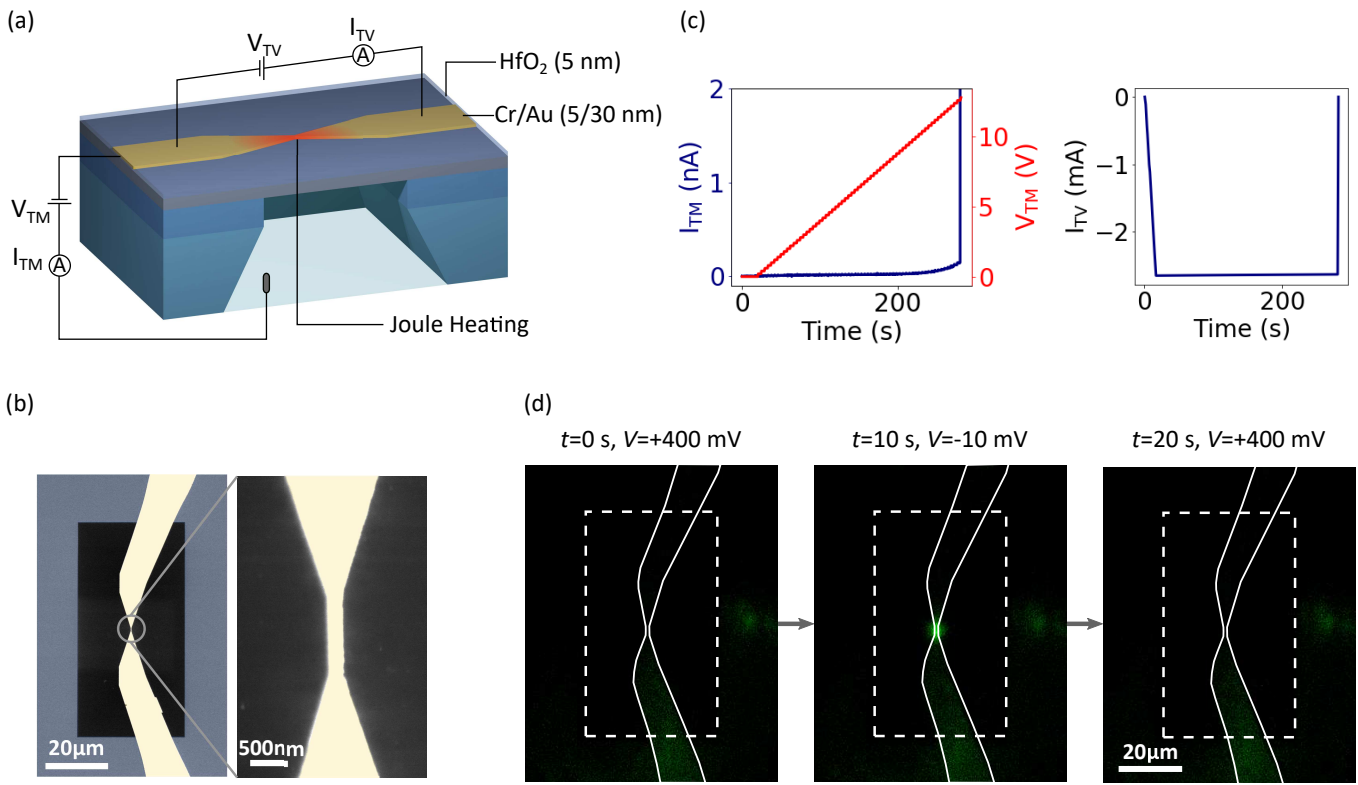}
    \caption{(a) Schematic of the device geometry and the experimental setup used to fabricate nanopores self-aligned with an on-chip metal nanoconstriction. The device layers are the same as labeled in Fig.~\ref{findpores}(a). (b) False-colour scanning electron micrographs of the electrode geometry. The left micrograph shows the electrode geometry over the suspended region of SiN$_x$ while the right micrograph shows a higher magnification micrograph at the centre of the metal nanoconstriction. (c) Example traces of the transverse current measured across the metal nanoconstriction (\itv) and the transmembrane current measured across the membrane (\itm) during CBD. The maximum voltage applied across the metal nanoconstriction (\vtv) is 1.2$\,$V. (d) Fluorescence images of the nanopore position after CBD. The three micrographs represent an image before (left), during (middle), and after (right) the application of a voltage that drives Ca$^{2+}$ ions through the nanopore. The dashed white box shows the region of suspended SiN$_x$ while the solid white lines show the position of the on-chip electrodes.}
    \label{flocal}
\end{figure*}

To test this hypothesis, we fabricated metal nanoconstrictions on SiN$_x$ membranes. We choose to use this device geometry since passing a current through a metal nanoconstriction results in localised Joule heating at the center of the constriction as the current density is highest here. A schematic of this device geometry and experimental setup is shown in Fig.~\ref{flocal}(a). False-colour SEM images of the electrodes over the suspended region of SiN$_x$ is shown in Fig.~\ref{flocal}(b). Aside from the different electrode geometry, the device structure is similar to that shown in Fig.~\ref{findpores}(a) with the exception of the electrode material (Cr/Au 5/30$\,$nm) and the passivation layer (5$\,$nm HfO$_2$). A detailed description of the device geometry and the fabrication procedure for these devices is provided in SI 1.

Previous studies have extensively utilised metal nanoconstrictions to create nanogap electrodes via electromigration by passing a large current density through the constriction \cite{Tsutsui2011,Gehring2019,Perrin2015}. Electromigration is a process whereby energy from electrons is transferred to the metal atoms resulting in their migration and the formation of a nanogap \cite{Hadeed2007}. This process is aided by Joule heating at the center of the constriction which increases the mobility of the metal atoms in this region \cite{Jeong2014}. Indeed, previous studies have demonstrated that the temperature at the center of a gold constriction can reach up to $\sim$390$\,^{\circ}$C prior to electromigration \cite{Jeong2014} (the exact value will likely vary depending on the geometry of the electrodes). To understand the operating limits of our devices, we increased the voltage applied across the nanoconstriction until a drop in the current was observed indicating the formation of a nanogap via electromigration (SI 9). This typically occurs at voltage of $\sim$1.5$\,$V and a corresponding current of $\sim$3$\,$mA for our device geometry. Assuming a width of of the constriction of 200$\,$nm and a thickness of 30$\,$nm, the current density at the narrowest part of the constriction immediately prior to electromigration is approximately 5x10$^{11}$A/m$^2$. 

\begin{figure}
    \centering
    \includegraphics[width=7cm]{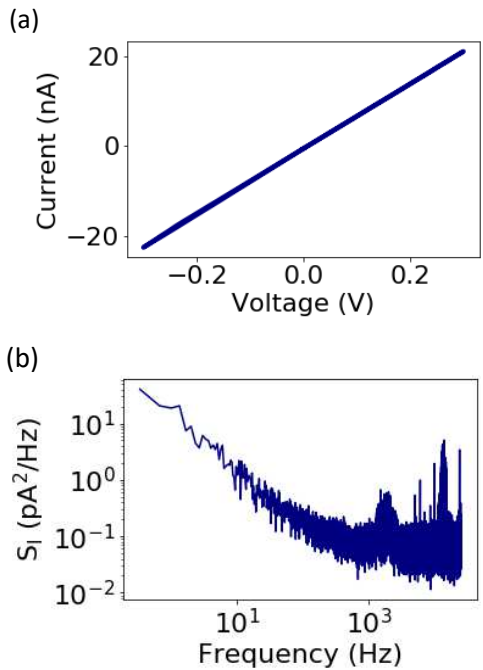}
    \caption{(a) Ionic current through a nanopore as a function of the transmembrane potential for a nanopore created via the CBD protocol described in Sec \ref{sindpores}. (b) Noise spectrum in the ionic current through the created nanopore. Measurement was performed at 200$\,$mV transmembrane potential. Measurements were performed by applying the voltage to the Ag/AgCl electrode in the $trans$ chamber while holding the Ag/AgCl electrodes in the $cis$ chamber and the on-chip electrodes at ground.}
    \label{fcharpore}
\end{figure}

\begin{figure*}
    \centering
    \includegraphics[width=\textwidth]{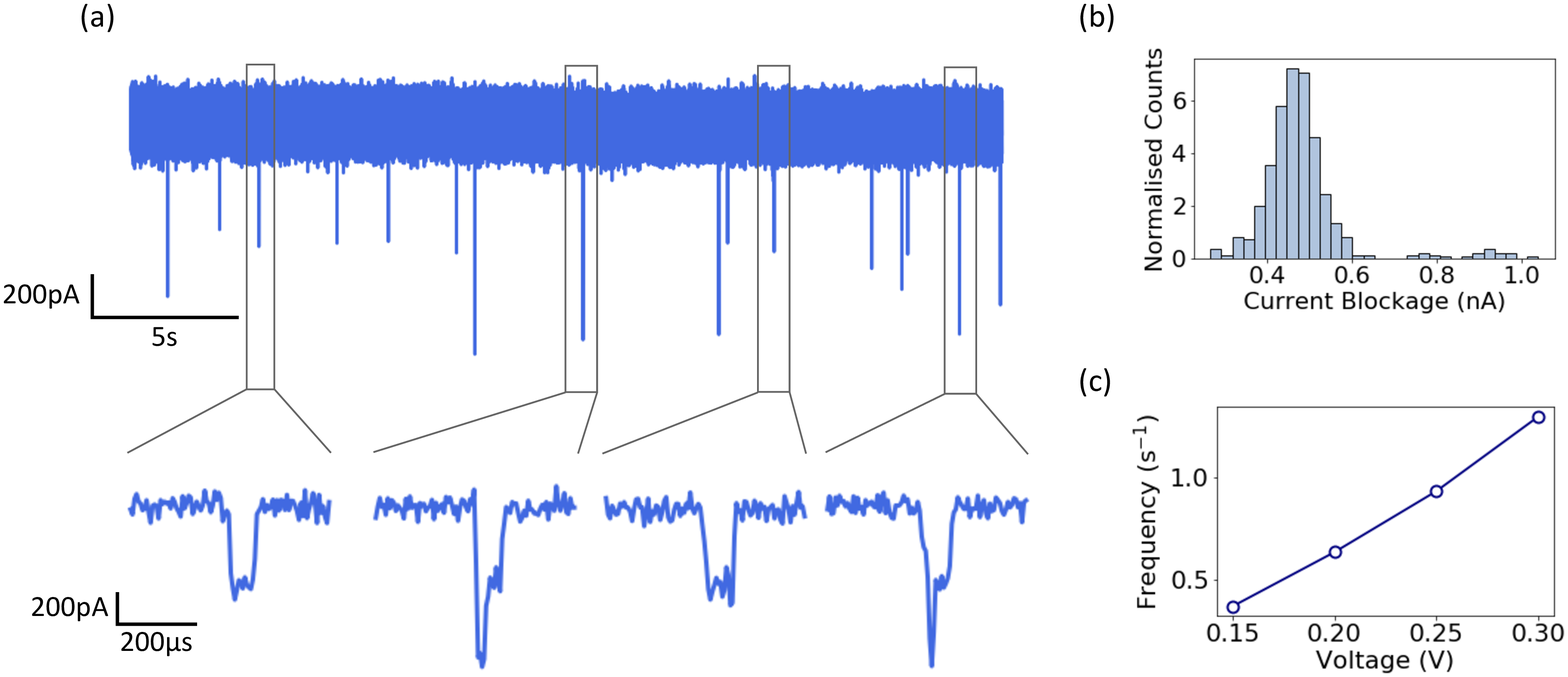}
    \caption{(a) Example current-time trace demonstrating translocation of 1$\,$kbp dsDNA through a nanopore created via the procedure described in Sec.~\ref{sindpores}. The lower panels show examples of several events demonstrating the translocation of folded and unfolded DNA. (b) A histrogram of the blockade levels for the extracted translocation events. (c) Translocation frequency as a function of the applied voltage. All experiments were performed by applying the voltage to the Ag/AgCl electrode in the $trans$ chamber while holding the Ag/AgCl in the $cis$ chamber and the on-chip electrodes at ground.}
    \label{fsensing}
\end{figure*}

To perform CBD on these devices, a transverse voltage (\vtv) was applied across the metal nanoconstriction while measuring the resulting current (\itv). \vtv{} was limited to 1.2$\,$V to ensure significant Joule heating while avoiding electromigration of the nanoconstriction. Simultaneous to this, an increasing transmembrane voltage (\vtm) was applied between the metal nanoconstriction and the electrolyte solution while measuring the resulting current (\itm). When a spike in \itm{} was observed indicating nanopore formation both \vtm{} and \vtv{} were reduced to zero. Figure \ref{flocal}(c) shows typical CBD measurements while simultaneously heating the membrane by passing a large current density through the metal nanoconstriction. To determine the position of the resulting nanopore, fluorescence imaging was performed. As shown in Fig.~\ref{flocal}(d), the nanopore formed at the narrowest part of the constriction. This is consistent with the fact that the Joule heating is highest in this region. This result was reproducible across multiple devices (SI 10). SEM images of the devices following breakdown also confirm that the nanopore was formed at the narrowest part of the nanoconstriction (SI 11). 

It is noted that following breakdown, the current measured across the metal nanoconstriction dropped to approximately zero [Fig.~\ref{flocal}(c)]. This suggests that damage has occurred to the metal nanoconstriction during breakdown such that the two sides of the constriction are no longer electrically connected. This was confirmed from SEM images of the device after nanopore formation (SI 11). Such damage may be due to the significant heating which increases the conductivity of the electrolyte leading to rapid nanopore expansion following breakdown \cite{Leung2020}. Indeed, nanopores created using the above protocol were typically larger than $\sim$50$\,$nm as observed from SEM images and confirmed by measuring the ionic current through the pore following breakdown. Due to limits in the electronics used in this work, a relatively slow feedback frequency was used to reduce the voltage after breakdown (the voltage is reduced up to $\sim$70$\,$ms following breakdown). It is likely that employing faster feedback conditions will enable the creation of smaller nanopores with minimal damage to the surrounding nanoelectrodes. Changing the device geometry such as (i) reducing the thicknesses of the SiN$_x$ membrane, (ii) modifying the electrode geometry or material, and (iii) changing the thickness or composition of the passivation layer may also enable the fabrication of smaller nanopores. This is an area that should be explored further in future studies.

\section{Biomolecular Sensing}

To demonstrate the utility of the nanopores created using the CBD strategy introduced here, we have performed biomolecular sensing using these devices. To do this, a single nanopore was created using the CBD protocol described in Sec.~\ref{sindpores}. Following breakdown, reservoirs on both sides of the membrane were filled with 3.6$\,$M LiCl and the device was left overnight for the ionic current through the nanopore to stabilise \cite{Briggs2014,Pud2015}. To perform ionic current measurements, a voltage was applied to the Ag/AgCl electrode in the $trans$ chamber while the Ag/AgCl electrode in the $cis$ chamber and the on-chip electrodes were held at 0$\,$V (ground). As shown in Fig.~\ref{fcharpore}(a), a linear ionic current as a function of the applied voltage was measured. The size of the nanopore estimated from the conductance is $\sim$13$\,$nm \cite{Kowalczyk2011}. The normalised noise spectrum of the nanopore at an applied voltage of +200$\,$mV is shown in Fig.~\ref{fcharpore}(b). The noise spectrum is comparable to that observed in solid-state nanopores created via TEM drilling \cite{Fragasso2020} or CBD when the breakdown voltage is applied via electrolyte solutions in contact with both sides of the membrane \cite{Waugh2020}. 

To demonstrate the ability to perform single-molecule biosensing using these nanopores, 1$\,$kbp dsDNA (NoLimits DNA, ThermoFisher) was added to the $cis$ chamber resulting in a DNA concentration of $\sim$5$\,$nM in that reservoir. As shown in Fig.~\ref{fsensing}(a), transient drops in the ionic current were observed upon application of a positive voltage which drives DNA through the pore. These transient drops were not observed prior to DNA being injected into the $cis$ chamber, or when a negative voltage was applied to the $trans$ chamber. Upon closer inspection of the current transients, one can observe both single-level and two-level drops in the current [bottom panel of Fig.~\ref{fsensing}(a)]. This can also be observed from the histogram of the current change during the extracted translocation events [Fig.~\ref{fsensing}(b)]. Observing two levels in the current blockade is typical of dsDNA translocation through a solid-state nanopore with a diameter at least twice as large as the diameter of dsDNA ($\sim$2.3$\,$nm) as is the case in this work \cite{Chen2004a}. The two-level current blockades correspond to the translocation of DNA in a folded manner through the nanopore while the single-level current blockades correspond to the translocation of DNA in a unfolded manner \cite{Chen2004a}. A linear increase in the translocation frequency as a function of applied voltage is also observed [Fig.~\ref{fsensing}(c)]. This indicates that the translocation frequency is primarily limited by diffusion of the analyte to the capture radius of the pore as expected for our nanopore geometry \cite{Chen2004a,Grosberg2010}. These results confirm that nanopores fabricated via the CBD strategy introduced here can be used for single-molecule biosensing applications.

\section{Conclusion and Outlook}

Here we have introduced a new CBD strategy to fabricate solid-state nanopores whereby breakdown is performed by applying a voltage between an on-chip electrode and an electrolyte solution in contact with the other side of the membrane. Since the applied electric field is localised to the on-chip electrode, this technique enables us to independently fabricate and localise multiple nanopores in the membrane. Moreover, by using the on-chip electrodes to locally heat the membrane we are able to create nanopores that are self-aligned with on-chip nanostructures. This was demonstrated by performing CBD simultaneous to passing a large current density through a metal nanoconstriction to induce localised Joule heating. Here, it was shown that the fabricated nanopore forms at the narrowest part of the constriction where Joule heating is highest. We believe this is a versatile technique that could be used to integrate nanopores with a range of on-chip nanostructures such as field-effect sensors and nanogap electrodes. In SI 12 we describe some possible geometries and voltage protocols that could be explored in future studies to self-align nanopores with various on-chip nanostructures via this technique. 

CBD is quickly becoming one of the most popular techniques to fabricate solid-state nanopores. This technique is particularly appealing as it does not require expensive equipment and can be fully automated thus making it accessible to the wider research community. Furthermore, CBD creates nanopores in the electrolyte environment which measurements are performed in which can reduce problems associated with wetting of the nanopore. Our new CBD strategy extends this popular nanopore fabrication technique for applications where it is desirable to fabricate arrays of nanopores integrated with on-chip nanostructures. We believe that the further development of our CBD strategy can significantly accelerate research into the development of solid-state nanopores integrated with complementary nanostructures and open up routes for the commercial development of these devices. 

\section{Conflicts of Interest}

The University of Oxford has filed a patent application for the technologies described here for which J. P. F., J. L. S., and J. A. M. have an interest in. J. R. Y. is a principal in Nanopore Solutions whose fluidic devices were used in this study. 

\section{Acknowledgments}

The authors would like to thank Kevin Lester and Daryl Briggs for their assistance with preparing the wafers. J. P. F. thanks the Oxford Australia Scholarship committee and the University of Western Australia for Funding. Substrate, membrane and some electrode fabrication was conducted at the Center for Nanophase Materials Sciences, which is a DOE Office of Science User Facility. J. R. Y. was funded by an FCT contract according to DL57/2016, [SFRH/BPD/80071/2011]. Work in J. R. Y.’s lab was funded by national funds through FCT - Fundação para a Ciência e a Tecnologia, I. P., Project MOSTMICRO-ITQB with refs UIDB/04612/2020 and UIDP/04612/2020 and Project PTDC/NAN-MAT/31100/2017. J. M. was supported through the UKRI Future Leaders Fellowship, Grant No. MR/S032541/1, with in-kind support from the Royal Academy of Engineering. A. P. I. and J. B. E. acknowledge support from BBSRC grant BB/R022429/1, EPSCR grant EP/P011985/1, and Analytical Chemistry Trust Fund grant 600322/05. This project has also received funding from the European Research Council (ERC) under the European Union's Horizon 2020 research and innovation programme (grant agreement No 724300 and 875525). O. D. and STEM investigations were supported by the Center for Nanophase Materials Sciences (CNMS), a U.S. Department of Energy, Office of Science User Facility. The authors would like to thank Andrew Briggs for providing funding for some of the equipment and facilities used.

\end{document}